# Playing Mastermind With Constant-Size Memory

Benjamin Doerr and Carola Winzen*

Max-Planck-Institut für Informatik, Saarbrücken, Germany
{doerr|winzen}@mpi-inf.mpg.de

## Abstract

We analyze the classic board game of Mastermind with $n$ holes and a constant number of colors. A result of Chvátal (Combinatorica 3 (1983), 325-329) states that the codebreaker can find the secret code with $\Theta(n/\log n)$ questions. We show that this bound remains valid if the codebreaker may only store a constant number of guesses and answers. In addition to an intrinsic interest in this question, our result also disproves a conjecture of Droste, Jansen, and Wegener (Theory of Computing Systems 39 (2006), 525-544) on the memory-restricted black-box complexity of the OneMax function class.

## 1 Introduction

The original *Mastermind* game is a board game for two players invented in the seventies by Meirowitz. It has pegs of six different colors. The goal of the *codebreaker*, for brevity called *Paul* here, is to find a color combination made up by *codemaker* (called *Carole* in the following). He does so by guessing color combinations and receiving information on how close this guess is to Carole's secret code. Paul's aim is to use as few guesses as possible.

For a more precise description, let us call the colors 1 to 6. Write $[n] := \{1, \ldots, n\}$ for any $n \in \mathbb{N}$. Carole's secret code is a length-4 string of colors, that is, a $z \in [6]^4$. In each iteration, Paul guesses a string $x \in [6]^4$ and Carole replies with a pair $(\text{eq}(z, x), \pi(z, x))$ of numbers. The first number, $\text{eq}(z, x)$, which is usually indicated via black answer-pegs, is the number of positions, in which Paul's and Carole's string coincide. The other number, $\pi(z, x)$, usually indicated by white answer-pegs, is the number of additional pegs having the right color, but being in the wrong position. Formally $\text{eq}(z, x) := |\{i \in [4] \mid z_i = x_i\}|$ and $\pi(z, x) := \max_{\rho \in S_4} |\{i \in [4] \mid z_i = x_{\rho(i)}\}| - \text{eq}(z, x)$, where $S_4$ denotes the set of all permutations of $[4]$. Paul "wins" the game if he guesses Carole's string, that is, if Carole's answer is $(4, 0)$.

We are interested in strategies for Paul that guarantee him to find the secret code with few questions. We thus adopt a worst-case view with respect to Carole's secret code. This is equivalent to assuming that Carole may change her hidden string at any time as long as it remains consistent with all previous answers (*devil's strategy*).

**Previous Results.** Mathematics and computer science literature produces a plethora of results on the Mastermind problem. For the original game with 6 colors and 4 positions, Knuth [Knu77] showed that Paul needs at most four queries until being able to identify Carole's string (which he may query in the fifth iteration to win the game).

---

*Carola Winzen is a recipient of the Google Europe Fellowship in Randomized Algorithms. This research is supported in part by this Google Fellowship.

Chvátal [Chv83] studies a general version of this game with $k$ colors and $n$ positions, that is, the secret code is a length-$n$ string $z \in [k]^n$. Denote by $d(n,k)$ the minimum number of guesses that enable Paul to win the game for any secret code. Chvátal proves that for $k < n^{1-\varepsilon}$, $\varepsilon > 0$ an arbitrarily small constant, we have $d(n,k) = O(\frac{n \log k}{\log n - \log k})$. More precisely, he shows that for any $\varepsilon > 0$ and $n$ sufficiently large, $(2+\varepsilon)\frac{n(1+2\log k)}{\log n - \log k}$ guesses chosen from $[k]^n$ independently and uniformly at random, with high probability, suffice to distinguish between all possible codes (that is, each secret code leads to a different sequence of answers). Therefore, the secret code can be determined after that many guesses. This remains true if Carole replies only with black answer-pegs, that is, if for any of Paul's guesses $x$ she reveals to him only eq$(z,x)$, the number of bits, in which her and Paul's string coincide.

For larger values of $k$, the following is known. For $n \leq k \leq n^2$, Chvátal proves $d(n,k) \leq 2n\log k + 4n$ and for $k = \omega(n^2 \log n)$ he shows $(k-1)/n \leq d(n,k) \leq \lceil k/n \rceil + d(n, n^2)$. These results have subsequently been improved. Chen, Cunha, and Homer [CCH96] show that $d(n,k) \leq 2n\lceil \log n \rceil + 2n + \lceil k/n \rceil + 2$ for $k \geq n$. Goodrich [Goo09] proves $d(n,k) \leq n\lceil \log k \rceil + \lceil (2-1/k)n \rceil + k$ for arbitrary $k$.

For $k = 2$ colors, the Mastermind problem is related to the well-studied coin weighing problem. For this reason, first results on this problem date back to years as early as 1963, when Erdős and Rényi [ER63] show that $d(n,2) = \Theta(n/\log n)$.

Concerning the computational complexity, Stuckman and Zhang [SZ06] show that it is $NP$-hard to decide whether a given sequence $(x^{(i)}, (\text{eq}^{(i)}, \pi^{(i)}))_{i=1}^t$ of queries $x^{(i)}$ and answers $(\text{eq}^{(i)}, \pi^{(i)})$ of black and white pegs has a secret code leading to these answers, i.e., whether there exists a string $z \in [k]^n$ such that $\text{eq}(z, x^{(i)}) = \text{eq}^{(i)}$ and $\pi(z, x^{(i)}) = \pi^{(i)}$ for all $i \in [t]$. Goodrich [Goo09] proves that this is already $NP$-hard if we only ask for consistence with the black answer-peg replies eq$^{(i)}$.

**Our results.** Originally motivated by a conjecture on black-box complexities (cf. Section 2), we study a memory-restricted version of the Mastermind problem. Since this original motivation asks for the case of two colors only, we restrict ourselves to the number $k$ of colors being constant, though clearly our methods can also be used to analyze larger numbers of colors.

The memory-restriction can be briefly described as follows. Given a memory of size $m \in \mathbb{N}$, Paul can store up to $m$ guesses and Carole's corresponding replies. Based *only* on this information, Paul decides on his next guess. After receiving Carole's reply, based only on the content of the memory, the current guess, and the current answer, he decides which $m$ out of the $m+1$ strings and answers he keeps in the memory. Note that our memory restriction means that Paul truly has no other memory, in particular, no iteration counters, no experience that certain colors are not used, and so one. So formally Paul's strategy consists of a guessing strategy, which can be fully described by a mapping from $m$-sets of guesses and answers to strings $x \in [k]^n$, and a forgetting strategy, which maps $(m+1)$-sets of guesses and answers to $m$-subsets thereof.

Clearly, a memory-restriction makes Paul's life not easier. The $O(n/\log n)$ strategies by Erdős and Rényi [ER63] and by Chvátal [Chv83] do use the full history of guesses and answers and thus only work with a memory of size $\Theta(n/\log n)$. Surprisingly, this amount of memory is not necessary. In fact, one single memory cell is sufficient.

**Theorem 1.** *Let $k \in \mathbb{N}$. For all $n \in \mathbb{N}$, Paul has a size-one memory strategy winning the Mastermind game with $k$ colors and $n$ positions in $O(n/\log n)$ guesses. This remains true if we allow*

Carole to play a devil's strategy and if Carole only reveals the number of fully correct pegs eq$(x, z)$ ("black answer-peg version of Mastermind").

The bound in Theorem 1 is asymptotically tight. A lower bound of $\Omega(n/\log n)$ is already true without memory restrictions. This follows easily from an information theoretic argument, cf. [ER63] or [Chv83]. Our result disproves a conjecture of Droste, Jansen, and Wegener [DJW06], who believed that a lower bound of $\Omega(n \log n)$ should hold for the 2-color black answer-peg Mastermind problem with memory-restriction one.

The proof of Theorem 1 is quite technical. For a clearer presentation of the ideas, we first consider the size-two memory-restricted model, cf. Section 3. The proof of Theorem 1 is given in Section 4. Before going into the proofs, in the following section we sketch the connection between Mastermind games and black-box complexities.

## 2 Mastermind and Black-Box Complexities

In this section, we describe the connection between the Mastermind game and black-box complexity. The reader only interested in the Mastermind result may skip this section without loss.

Roughly speaking, the *black-box complexity* of a set of functions is the number of function evaluations needed to find the optimum of an unknown member from that set. Since problem-unspecific search heuristics such as randomized hill-climbers, evolutionary algorithms, simulated annealing etc. do optimize by repeatedly generating new search points and evaluating their objective values ("fitness"), the black-box complexity is a lower bound on the efficiency of such general-purpose heuristics [DJW06].

**Black-Box Complexity.** Let $\mathcal{S}$ be a finite set. A (randomized) algorithm following the scheme of Algorithm 1 is called black-box optimization algorithm for functions $\mathcal{S} \to \mathbb{R}$.

---

**Algorithm 1:** Scheme of a black-box algorithm for optimizing $f : \mathcal{S} \to \mathbb{R}$

1 **Initialization:** Sample $x^{(0)}$ according to some probability distribution $p^{(0)}$ on $\mathcal{S}$;
2 Query $f(x^{(0)})$;
3 **for** $t = 1, 2, 3, \ldots$ **do**
4 $\quad$ Depending on $\big((x^{(0)}, f(x^{(0)})), \ldots, (x^{(t-1)}, f(x^{(t-1)}))\big)$ choose a probability distribution $p^{(t)}$ on $\mathcal{S}$ and sample $x^{(t)}$ according to $p^{(t)}$;
5 $\quad$ Query $f(x^{(t)})$;

---

For such an algorithm $A$ and a function $f : \mathcal{S} \to \mathbb{R}$, let $T(A, f) \in \mathbb{R} \cup \{\infty\}$ be the expected number of fitness evaluations until $A$ queries for the first time some $x \in \arg\max f$. We call $T(A, f)$ the *runtime of $A$ for $f$*. For a class $\mathcal{F}$ of functions $\mathcal{S} \to \mathbb{R}$, the *$A$-black-box complexity of $\mathcal{F}$* is $T(A, \mathcal{F}) := \sup_{f \in \mathcal{F}} T(A, f)$, the worst-case runtime of $A$ on $\mathcal{F}$. Let $\mathcal{A}$ be a class of black-box algorithms for functions $\mathcal{S} \to \mathbb{R}$. Then the *$\mathcal{A}$-black-box complexity of $\mathcal{F}$* is $T(\mathcal{A}, \mathcal{F}) := \inf_{A \in \mathcal{A}} T(A, \mathcal{F})$. If $\mathcal{A}$ is the class of all black-box algorithms, we also call $T(\mathcal{A}, \mathcal{F})$ the *unrestricted black-box complexity* of $\mathcal{F}$.

As said, the unrestricted black-box complexity is a lower bound for the efficiency of randomized search heuristics optimizing $\mathcal{F}$. Unfortunately, often this lower bound is not very useful. For example, Droste, Jansen, and Wegener [DJW06] observe that the $NP$-complete MaxClique problem on graphs of $n$ vertices has a black-box complexity of only $O(n^2)$.

**Black-Box Algorithms with Bounded Memory.** As a possible solution to this dilemma, Droste, Jansen, and Wegener suggest to restrict the class of algorithms considered from all black-box optimization algorithms to a reasonably large subset. A natural restriction is to forbid the algorithm to exploit the whole history of search points evaluated. This is motivated by the fact that many heuristics, e.g., evolutionary algorithms, only store a bounded size *population* of search points. Simple hill-climbers or the Metropolis algorithm even store only one single search point.

Algorithm 2 is the scheme of a black-box algorithm with bounded memory of size $\mu$. It is important to note that a black-box algorithm with bounded memory is not allowed to access any other information than the one stored in the $\mu$ pairs $(x^{(1)}, f(x^{(1)})), \ldots, (x^{(\mu)}, f(x^{(\mu)}))$, which are currently stored in the memory and, in the selection step, also the information provided by $(x^{(\mu+1)}, f(x^{(\mu+1)}))$. In particular, the algorithm does not have access to an iteration counter.

---

**Algorithm 2:** Scheme of a black-box algorithm with memory of size $\mu$ for optimizing function $f : \mathcal{S} \to \mathbb{R}$

1 **Initialization:** $\mathcal{M} \leftarrow \emptyset$;
2 **for** $t = 1, 2, \ldots$ **do**
3   Depending (only) on $\mathcal{M}$ choose a probability distribution $p$ on $\mathcal{S}$ and sample $x^{(\mu+1)}$ according to $p$; //variation step
4   Query $f(x^{(\mu+1)})$;
5   Select $\mathcal{M} \subseteq \mathcal{M} \cup \{(x^{(\mu+1)}, f(x^{(\mu+1)}))\}$ of size $|\mathcal{M}| \leq \mu$; //selection step

---

**Mastermind and the OneMax function class.** A test function often regarded to analyze how the randomized search heuristic under investigation progresses in easy parts of the search space, is the simple ONEMAX function ONEMAX : $\{0,1\}^n \to \mathbb{R}, x \mapsto \sum_{i=1}^{n} x_i$. Note that ONEMAX$(x) =$ eq$((1,\ldots,1), x)$ for all $x \in \{0,1\}^n$. In fact, for any $z \in \{0,1\}^n$, eq$(z, \cdot)$ yields an equivalent optimization problem. Let us denote by ONEMAX$_n := \{$eq$(z, \cdot) \mid z \in \{0,1\}^n\}$ the class of all these functions.

Due to a coupon collector effect, many classical randomized search heuristics like randomized local search or the $(\mu + \lambda)$ evolutionary algorithm (with $\mu, \lambda$ constants) need $\Theta(n \log n)$ function evaluations to optimize ONEMAX$_n$.

As a moments thought reveals, black-box algorithms optimizing ONEMAX$_n$ correspond to strategies for Paul in the Mastermind game (without memory restriction) with two colors and only black answer-pegs used. Hence the unrestricted black-box complexity of ONEMAX$_n$ is $\Theta(n/\log n)$ by the results of Erdős and Rényi [ER63] and Chvátal [Chv83].

This connection was seemingly overlooked so far in the randomized search heuristics community, where Droste, Jansen, and Wegener [DJW06] prove an upper bound of $O(n)$ and later Anil and Wiegand [AW09] prove the asymptotically correct bound of $O(n/\log n)$. Since already the first bound is lower than what many randomized search heuristics achieve, Droste, Jansen, and Wegener suggest to investigate the memory-restricted black-box complexity of ONEMAX$_n$. They conjecture in [DJW06, Section 6] that a memory restriction of size one leads to a black-box complexity of order $\Theta(n \log n)$.

Again, clearly, the memory-restricted black-box complexity of ONEMAX$_n$ and optimal strategies for Mastermind with two colors, black answer-pegs only, and a corresponding memory restriction are equivalent questions. Consequently, our result can be rephrased to saying that the black-box complexity of ONEMAX$_n$ even with the memory restricted to one is $\Theta(n/\log n)$, disproving the conjecture of Droste, Jansen, and Wegener.

## 3 The Mastermind Game with Memory of Size Two

Since the proof of Theorem 1 is quite technical, we give in this section a simpler proof showing that with a memory of size two Paul can win the game using only $O(n/\log n)$ guesses. Already this proof contains many ingredients needed to prove Theorem 1, e.g., the use of the random guessing strategy with limited memory, the block-wise determination of the secret code, and the simulation of iteration counters in the memory.

Let $k \geq 2$ be the number of colors used. In particular for $k = 2$, it will be convenient to label the colors from 0 to $k-1$. Let us denote the set of colors by $\mathcal{C} := [0..k-1] := \{0, 1, \ldots, k-1\}$. We assume that $k$ is a constant and that the number $n$ of positions in the string is large, that is, all asymptotic notation is with respect to $n$.

**Theorem 2.** *Paul has a size-two memory strategy winning the black answer-peg only Mastermind game with $k$ colors and $n$ positions in $O(n/\log n)$ guesses. This remains true if we allow Carole to play a devil's strategy.*

As many previous works, the proof of Theorem 2 heavily relies on *random guessing*. For the case of $k = 2$ colors, already Erdős and Rényi [ER63] showed that there is a $t \in \Theta(n/\log n)$ such that $t$ guesses $x^{(1)}, \ldots, x^{(t)}$ chosen from $\{0,1\}^n$ independently and uniformly at random, together with Carole's black answer-peg answers, uniquely define the hidden code. This was generalized by Chvátal [Chv83] to the following result.

**Theorem 3** (from [Chv83]). *Let $\varepsilon > 0$, let $n > n(\varepsilon)$ be sufficiently large and let $k < n^{1-\varepsilon}$. Let $x^{(1)}, \ldots, x^{(t)}$ be $t \geq (2+\varepsilon)\frac{n(1+2\log k)}{\log n - \log k}$ samples chosen from $\mathcal{C}^n$ independently and uniformly at random. Then for all $z \in \mathcal{C}^n$, the set*

$$\mathcal{S}^{consistent} := \{y \in \mathcal{C}^n \mid \forall i \in [t] : \mathrm{eq}(y, x^{(i)}) = \mathrm{eq}(z, x^{(i)})\}$$

*satisfies* $\mathrm{E}[|\mathcal{S}^{consistent}|] \leq 1 + 1/n$.

Since the strategy implicit in Theorem 3 needs a memory of size $\Theta(n/\log n)$, we cannot apply it directly in our setting. We can, however, adapt it to work on smaller portions ("blocks") of the secret code, and this with much less memory.

Let $y \in \mathcal{C}^n$ and let $B \subseteq [n]$ be a block (i.e., an interval) of size $s := \lceil\sqrt{n}\rceil$. As we shall see, by $t \in O(s/\log s)$ times guessing a string obtained from $y$ by replacing the colors in $B$ by randomly chosen ones (and guessing $k$ additional *reference strings*), we can determine $z_{|B}$, the part of the secret code $z$ in block $B$.

We can do so with a memory of size two only. We store the string obtained from $y$ by altering it on $B$ (*sampling string*) in one cell. Note that we do not need to remember $y$, as we only need to ensure that our guesses agree in the positions $[n] \setminus B$. We use the other memory cell (*storage string*, in the following typically denoted by $x$) to store the random substrings of length $s$ substituted into $y$ at $B$, and Carole's answers. Note that each such answer can be encoded in binary using $\ell_n \in O(\log n)$ entries of the string. Hence the $t$ guesses and answers can be memorized using a total number of $t(s + \ell_n) = O(n/\log n)$ positions.

**Algorithm 3:** An almost size-two memory-restricted algorithm winning the $k$-color black answer-peg only Mastermind game in $O(n/\log n)$ guesses. **Remark:** $x$ denotes the unique string in $\mathcal{M}$ with $x_n = 1$ and $y$ denotes the unique string in $\mathcal{M}$ with $y_n = 0$.

1 **Initialization:** $y \leftarrow [0\ldots 0]$;
2    Query $\text{eq}(z, y)$ and update $\mathcal{M} \leftarrow \{(y, \text{eq}(z, y))\}$;
3 **for** $i = 1$ **to** $\lceil (n-1)/s \rceil$ **do**
4     $x \leftarrow [0\ldots 0|1]$; //initialization of $x$
5     Query $\text{eq}(z, x)$ and update $\mathcal{M}$ by adding ($i = 1$) or replacing ($i > 1$) $(x, \text{eq}(z, x))$ in $\mathcal{M}$;
6     **for** $q = 0$ **to** $t + k - 1$ **do**
7       **if** $q < k$ **then** $y \leftarrow \texttt{substitute}(y, B_i, [q\ldots q])$; //reference string
8       **else** $y \leftarrow \texttt{substitute}(y, B_i, r)$ where $r \in \mathcal{C}^{|B_i|}$ u.a.r.; //random guess
9       Query $\text{eq}(z, y)$ and update $\mathcal{M}$ by replacing $(y, \text{eq}(z, y))$;
10      $x \leftarrow [x_1 \ldots x_{p_1(x)} | \texttt{BLOCK}_i(y) | \texttt{binary}_{\ell_n}(\text{eq}(z, y)) | 1 | 0 \ldots 0 | 1]$; //add $y$'s info to $x$
11      Query $\text{eq}(z, x)$ and update $\mathcal{M}$ by replacing $(x, \text{eq}(z, x))$;
12     **while** $\Delta_i(y) < |B_i|$ **do**
13       $y \leftarrow \texttt{substitute}(y, B_i, w)$, where $w \in \mathcal{S}_i^{\text{consistent}}$ u.a.r.;
14       Query $\text{eq}(z, y)$ and update $\mathcal{M}$ by replacing $(y, \text{eq}(z, y))$;
15 **while** $\text{eq}(z, y) < n$ **do** $y \leftarrow \texttt{substitute}(y, \{n\}, c)$, where $c \in \mathcal{C}$ u.a.r., and query $\text{eq}(z, y)$;

This approach allows us to determine $s$ positions of $z$ using $t = O(s/\log s)$ guesses. Hence we can determine the secret code $z$ with $t\lceil n/s \rceil = O(n/\log n)$ guesses as desired.

In Algorithm 3 (notation used will be introduced below) we make this strategy more precise by giving it in pseudo-code. Note, however, that this algorithm does not fully satisfy the size-two memory restriction. The reason is that the queries do not only depend on the current state of the memory, but also on iteration counters and, e.g. in lines 9 and 11, on the program counter. Further below, in Algorithm 4 we shall remove this shortcoming with a few additional technicalities, which we are happy to spare for the moment.

Before we argue for the correctness of Algorithm 3, let us fix the notation. For any string $x \in \mathcal{C}^n$ we also write $x = [x_1 \ldots x_n]$. To ease reading, we allow ourselves to indicate different structural components of $x$ by vertical bars, e.g., $x = [x_1 \ldots x_p | x_{p+1} \ldots x_n]$. For $i \in [\lceil (n-1)/s \rceil]$ let $B_i := \{(i-1)s + 1, \ldots, is\} \cap [n-1]$, the positions of the $i$-th block. Set

$$\texttt{BLOCK}_i(x) := x_{|B_i} := [x_{(i-1)s+1} \ldots x_{\min\{is, n-1\}}],$$

the $i$-th block of $x$. For any string $r \in \mathcal{C}^{|B_i|}$ we define

$$\texttt{substitute}(x, B_i, r) := [x_1 \ldots x_{(i-1)s} | r | x_{\min\{is, n-1\}+1} \ldots x_n],$$

the string with the $i$-th block substituted by $r$. Similarly, let $\texttt{substitute}(y, \{n\}, c) := [y_1 \ldots y_{n-1} | c]$. Note that we do not assign the $n$-th position to any of the blocks. We do so because in Algorithms 3 and 4 we shall use that position to indicate, which one of the two strings in the memory $\mathcal{M}$ is the storage string (the unique $x \in \mathcal{M}$ with $x_n = 1$) and which one is the sampling string (the unique string $y \in \mathcal{M}$ with $y_n = 0$).

Let $p_1(x) := \max\{i \in [n-1] \mid x_i = 1\}$, the largest position $i < n$ of $x$ with entry "1". As mentioned above, we encode Carole's answers $\text{eq}(z, y) \in [0..n]$ in binary, using $\ell_n := \lceil \log n \rceil +$ 1 positions, and we denote this binary encoding of length $\ell_n$ by $\mathtt{binary}_{\ell_n}(\mathrm{eq}(z,y))$. By $\Delta_i(y)$ we denote the contribution of the $i$-th block to the value $\mathrm{eq}(z,y)$, i.e., $\mathrm{eq}(z,y)$ is the number of positions in the $i$-th block, in which Paul's guess $y$ and Carole's secret code $z$ coincide. Formally, $\Delta_i(y) := \mathrm{eq}(z_{|B_i}, y_{|B_i})$. Lastly, let $\mathcal{S}_i^{\mathrm{consistent}}$ be the set of strings $w$ of length $|B_i|$ such that $\mathtt{substitute}(z, B_i, w)$ is consistent with all of Carole's replies (formal definition follows). We shall see below that both $\Delta_i(y)$ and $\mathcal{S}_i^{\mathrm{consistent}}$ can be computed solely from the content of the memory cells (lines 12–14).

We now argue for the correctness of Algorithm 3. Let us consider the state of the memory after having sampled all $t$ random samples for the $i$-th block (that is, we are in lines 12–14). We show that based on the information given in the memory, we can restore the full history of guesses for the $i$-th block. To this end, first note that for any guess $y$ done in line 9, we used $s + \ell_n + 1$ positions in $x$ for storing its information (line 10; we add the additional "1" at the end to ease determining via $p_1(x)$ the positions in $x$, which have not yet been used for storing information). In lines 6–11 we first asked and stored $k$ non-random guesses $x^c = \mathtt{substitute}(y, B_i, [c \ldots c])$ and we stored these *reference strings* together with Carole's replies $\mathrm{eq}(z, x^c) = \sum_{h=1}^{\ell_n} 2^{h-1} x_{c(s+\ell_n+1)-h}$, $c \in [0..k-1]$. Therefore, for $j \in [t]$, the $j$-th random sample is $r^{(j)} = [x_{(k+j-1)(s+\ell_n+1)+1} \cdots x_{(k+j-1)(s+\ell_n+1)+|B_i|}]$ and the corresponding query was $y^{(j)} = \mathtt{substitute}(y, B_i, r^{(j)})$. We have stored Carole's reply to this guess in binary, and we can infer $\mathrm{eq}(z, y^{(j)}) = \sum_{h=1}^{\ell_n} 2^{h-1} x_{(k+j)(s+\ell_n+1)-h}$. This shows how to regain the full guessing history.

Next we show how to compute the contributions $\Delta_i(y^{(j)})$ of the entries in the $i$-th block. To this end, note that the constant substrings $[c \ldots c]$ in the reference strings $x^c$ in total contribute exactly $|B_i|$ to the sum $\mathrm{eq}(z, x^0) + \ldots + \mathrm{eq}(z, x^k)$. Formally, $\sum_{c=0}^{k-1} \mathrm{eq}([z_{(i-1)s+1} \cdots z_{\min\{is, n-1\}}], [c \ldots c]) = |B_i|$. Since all other positions of the sampling string $y$ are not changed during the phase, in which we determine the $i$-th block, we infer that

$$\Delta_i(y^{(j)}) = \mathrm{eq}(z, y^{(j)}) - (\mathrm{eq}(z, x^0) + \ldots + \mathrm{eq}(z, x^k) - |B_i|)/k \,.$$

Consequently, in lines 12–14, the algorithm can compute $\Delta_i(y^{(j)})$ for all $j \in [t]$. From this it can infer

$$\mathcal{S}_i^{\mathrm{consistent}} := \{\tilde{z} \in \mathcal{C}^{|B_i|} \mid \forall j \in [t] : \mathrm{eq}(\tilde{z}, \mathtt{BLOCK}_i(y^{(j)})) = \Delta_i(y^{(j)})\},$$

the set of possible code segments in $B_i$. By Theorem 3, the expected size of $\mathcal{S}_i^{\mathrm{consistent}}$ is bounded from above by $1 + 1/|B_i|$. Thus, in lines 12–14 we need an expected number of $1 + 1/|B_i|$ samples $w$ chosen from $\mathcal{S}_i^{\mathrm{consistent}}$ uniformly at random until we find a $y = \mathtt{substitute}(y, B_i, w)$ with $\Delta_i(y) = s$ (which implies that the $i$-th block of $y$ coincides with Carole's secret code). This shows how we determine the entries of the $i$-th block in an expected total number of $t = O(s/\log s)$ guesses.

When Algorithm 3 executes line 15, all but the last entry of $y$ coincide with Carole's secret code. Hence trying random colors in the $n$-th position finds the hidden code $z$ with an additional expected number of $k = \Theta(1)$ guesses.

To turn Algorithm 3 into a size-two memory-restricted one, we use the first $\ell_n$ entries of $x$ to store in binary the iteration counter $i$, which indicates the index of the block currently being under consideration. This will move the storage space for the guesses and answers by $\ell_n$ positions to the right. Formally, we define $i(x) := \sum_{h=0}^{\ell_n - 1} 2^h x_{\ell_n - h}$. The inner for loop needs no additional memory to be simulated, because we can learn from $p_1(x)$ how many guesses $q(x)$ have been queried already. More precisely, since storing each guess requires $s + \ell_n + 1$ positions and the first $\ell_n$ positions are used for indicating the number of already determined entries, we have $q(x) := (p_1(x) - \ell_n)/(s + \ell_n + 1)$.

**Algorithm 4:** A size-two memory-restricted algorithm winning the $k$-color black answer-peg only Mastermind game in $O(n/\log n)$ guesses. **Remark:** $x$ denotes the unique string in $\mathcal{M}$ with $x_n = 1$ and $y$ denotes the unique string in $\mathcal{M}$ with $y_n = 0$.

**1 Initialization:** Let $\mathcal{M} \leftarrow \emptyset$; // clear memory
**2 if** $\mathcal{M} = \emptyset$ **then**
**3** $\quad y \leftarrow [0...0]$; //first reference string
**4** $\quad$ Query $\text{eq}(z, y)$ and update $\mathcal{M} \leftarrow \{(y, \text{eq}(z, y))\}$;
**5 else if** $|\mathcal{M}| = 1$ **then**
**6** $\quad x \leftarrow [0...0|1]$; //initialization of storage string
**7** $\quad$ Query $\text{eq}(z, x)$ and update $\mathcal{M} \leftarrow \mathcal{M} \cup \{(x, \text{eq}(z, x))\}$;
**8 else if** $i(x) < \lceil (n-1)/s \rceil$ **then**
**9** $\quad$ **if** $x = [0\ldots 0|1]$ **or** $\Delta_{i(x)}(y) = |B_{i(x)}|$ **then**
**10** $\quad\quad x \leftarrow [\texttt{binary}_{\ell_n}(i(x)+1)|\texttt{BLOCK}_{i(x)+1}(y)|\texttt{binary}_{\ell_n}(\text{eq}(z,y))|1|0\ldots 0|1]$; //clear storage string and add first reference string
**11** $\quad\quad$ Query $\text{eq}(z, x)$ and update $\mathcal{M}$ by replacing $(x, \text{eq}(z, x))$;
**12** $\quad$ **else if** $Part(y, x) = 1$ **and** $q(x) < t + k$ **then**
**13** $\quad\quad$ **if** $q(x) < k$ **then** $y \leftarrow \texttt{substitute}(y, B_{i(x)}, [q(x)\ldots q(x)])$; //reference string
**14** $\quad\quad$ **else** $y \leftarrow \texttt{substitute}(y, B_{i(x)}, r)$ where $r \in \mathcal{C}^{|B_{i(x)}|}$ u.a.r.; //random guess
**15** $\quad\quad$ Query $\text{eq}(z, y)$ and update $\mathcal{M}$ by replacing $(y, \text{eq}(z, y))$;
**16** $\quad$ **else if** $Part(y, x) = 0$ **and** $\Delta_{i(x)}(y) < |B_{i(x)}|$ **then**
**17** $\quad\quad x \leftarrow [x_1 \ldots x_{p_1(x)}|\texttt{BLOCK}_{i(x)}(y)|\texttt{binary}_{\ell_n}(\text{eq}(z, y))|1|0\ldots 0|1]$; //add $y$'s info to $x$
**18** $\quad\quad$ Query $\text{eq}(z, x)$ and update $\mathcal{M}$ by replacing $(x, \text{eq}(z, x))$;
**19** $\quad$ **else if** $Part(y, x) = 1$ **and** $q(x) = t + k$ **then**
**20** $\quad\quad y \leftarrow \texttt{substitute}(y, B_{i(x)}, w)$ where $w \in \mathcal{S}^{\text{consistent}}_{i(x)}$ chosen u.a.r.;
**21** $\quad\quad$ Query $\text{eq}(z, y)$;
**22** $\quad\quad$ **if** $\Delta_{i(x)}(y) = |B_{i(x)}|$ **then** Update $\mathcal{M}$ by replacing $(y, \text{eq}(z, y))$;
**23 else if** $i(x) = \lceil (n-1)/s \rceil$ **then**
**24** $\quad y \leftarrow \texttt{substitute}(y, \{n\}, c)$ where $c \in \mathcal{C}\backslash\{y_n\}$ u.a.r.;
**25** $\quad$ Query $\text{eq}(z, y)$;
**26 Go to line 2;**

Lastly, we need to replace the sequential queries in lines 9 and 11 of Algorithm 3 (as this exploits information stored in the program counter). Fortunately, again we can deduce from the memory where we stand. We define a function $\texttt{Part}(y, x)$, which equals 1 if the information of $y$ has been added to the storage string $x$ already and which equals 0 otherwise. That is, we set

$$\texttt{Part}(y, x) = \begin{cases} 1, & \text{if } \sum_{i=1}^{\ell_n} 2^{i-1} x_{p_1(x)-i} = \text{eq}(z, y) \text{ and } \texttt{BLOCK}_{i(x)}(y) = [x_{p_1(x)-\ell_n-|B_{i(x)}|} \ldots x_{p_1(x)-\ell_n-1}] \\ 0, & \text{otherwise} \,. \end{cases}$$

Note that $\texttt{Part}(y, x) = 1$ indicates that the information of $y$ has been stored in $x$ also in the case that our current sample equals the previous one. This is no problem as then the current guess does not give any new information. Hence the use of $\texttt{Part}$ modifies the algorithm to sample $t$ random guesses without immediate repitition. Note that the probability to sample the same string $r \in \mathcal{C}^{|B_i(x)|}$ twice in a row is at most $1/2$ (if the last block consists only of one position and $k = 2$) and is typically much smaller. Hence, occurrences of this event have no influence on the asymptotic number of guesses needed to win the game.

With these modifications, Algorithm 3 becomes the truly size-two memory-restricted Algorithm 4.

## 4 Memory of Size One: Proof of Theorem 1

Compared to the situation in Section 3, Paul faces two additional challenges in the size-one memory-restricted setting. The obvious one is that he has less memory available, in particular, after a large part of the code has been determined and needs to be stored. The more subtle one is that he cannot any longer query a search point and then store whatever is worth storing in the second memory cell. With one memory cell, all he can do is to guess a new string and keep or forget it.

### 4.1 Linear Time Strategies

Before we give a proof of Theorem 1, let us discuss a linear time winning strategy, i.e., a strategy that allows Paul to find Carole's secret code in a linear expected number of guesses using one memory cell only. This linear time strategy will be used in the proof of Theorem 1 to determine the last $\Theta(n/\log n)$ entries of the secret code.

The basic idea of the linear time strategy is to test each position one by one, from left to right. Since we have just one memory cell, we need to indicate in this one string, which entries have been determined already. We do so by keeping all not yet determined entries at one identical value different from the one of the entry determined last. To this end, let us for all $x \in \mathcal{C}^n$ define

$$\operatorname{tn}(x) := \min\{i \in [n] \mid \forall j \in \{i, \ldots, n\} : x_j = x_i\},$$

the *tail number* of $x$. The following lemma describes the linear time strategy.

**Lemma 4.** *Let $x \in \mathcal{C}^n$. Furthermore, let us denote Carole's secret code by $z \in \mathcal{C}^n$. Let us assume that the first $\operatorname{tn}(x) - 1$ entries of $z$ have been determined (i.e., Carole can no longer change the entries of $[z_1 \ldots z_{\operatorname{tn}(x)-1}]$). Further assume that $x_i = z_i$ for all $i < \operatorname{tn}(x)$ and that $\mathcal{M} = \{(x, \operatorname{eq}(z, x))\}$ is the current content of the memory cell.*

*There is a size-one memory-restricted guessing procedure* `LinAlg` *that—even if Carole plays a devil's strategy—after an expected constant number of successive calls modifies the memory such that the string $y$ now in the memory satisfies $y_i = z_i$ for all $i \leq \operatorname{tn}(x)$ and $\operatorname{tn}(y) = \operatorname{tn}(x) + 1$. Every call of* `LinAlg` *requires only one guess.*

Interestingly, for the definition of `LinAlg`, we need to distinguish between the cases of $k = 2$ and $k \geq 3$ colors, as certain arguments exploit particular properties of these cases. For $k = 2$ colors, we have analyzed this strategy already in [DW11].

#### 4.1.1 The case of $k = 2$ colors $\mathcal{C} = \{0, 1\}$

In this section we prove that for $k = 2$ colors $\mathcal{C} = \{0, 1\}$, `LinAlg` is a procedure that requires, in expectation, three calls to modify the memory content by replacing the current string $x$ that is assumed to satisfy the conditions of Lemma 4, by a string $y$ with $\operatorname{tn}(y) = \operatorname{tn}(x) + 1$ and $y_i = z_i$ for all $i \leq \operatorname{tn}(x)$.

For all $i \in [n]$ let $e_i^n$ be the $i$-th unit vector of length $n$.

---

**Algorithm 5:** Routine `LinAlg` for $k = 2$ colors

1 **Assumption:** The string $x \in \{0,1\}^n$ in the memory satisfies $\operatorname{tn}(x) < n$ and $x_i = z_i$ for all $i < \operatorname{tn}(x)$;
2 Sample $y \in \{x \oplus e_{\operatorname{tn}(x)}^n, x \oplus \sum_{i=\operatorname{tn}(x)+1}^n e_i^n\}$ uniformly at random;
3 Query $\operatorname{eq}(z, y)$;
4 **if** $y = x \oplus e_{\operatorname{tn}(x)}^n$ **then**
5      **if** $\operatorname{eq}(z, y) > \operatorname{eq}(z, x)$ **then** $\mathcal{M} \leftarrow \{(y, \operatorname{eq}(z,y))\}$;
6 **else**
7      **if** $\operatorname{eq}(z, x) + \operatorname{eq}(z, y) = n + \operatorname{tn}(x)$ **then** $\mathcal{M} \leftarrow \{(y, \operatorname{eq}(z,y))\}$;

---

**Proposition 5** (from [DW11]). *For $k = 2$ colors, Algorithm 5 satisfies the conditions of Lemma 4. In expectation, three calls to routine* `LinAlg` *suffice.*

*Proof.* Let $x \in \{0,1\}^n$ be a bit string with $\operatorname{tn}(x) < n$ and $x_i = z_i$ for all $i < \operatorname{tn}(x)$.

Algorithm 5 samples with probability $1/2$ the string $y = x \oplus e_{\operatorname{tn}(x)}^n$, and with probability $1/2$ it samples $y = x \oplus \sum_{i=\operatorname{tn}(x)+1}^n e_i^n$. That is, either it flips only the $\operatorname{tn}(x)$-th bit of $x$ or it flips all "tail bits" but the tail numbered one.

If $y = x \oplus e_{\operatorname{tn}(x)}^n$, clearly we have $z_{\operatorname{tn}(x)} = y_{\operatorname{tn}(x)}$ if and only if $\operatorname{eq}(z,y) > \operatorname{eq}(z,x)$.

Therefore, let us assume that Algorithm 5 samples $y = x \oplus \sum_{i=\operatorname{tn}(x)+1}^n e_i^n$. We show that $z_{\operatorname{tn}(x)} = y_{\operatorname{tn}(x)} (= x_{\operatorname{tn}(x)})$ holds if and only if $\operatorname{eq}(z,x) + \operatorname{eq}(z,y) = n + \operatorname{tn}(x)$. By definition we have $y_i = x_i = z_i$ for all $i < \operatorname{tn}(x)$. Thus, the first $\operatorname{tn}(x) - 1$ bits of $x$ and $y$ contribute $2(\operatorname{tn}(x) - 1)$ to the sum $\operatorname{eq}(z,x) + \operatorname{eq}(z,y)$; formally,

$$\operatorname{eq}([z_1 \ldots z_{\operatorname{tn}(x)-1}], [x_1 \ldots x_{\operatorname{tn}(x)-1}]) + \operatorname{eq}([z_1 \ldots z_{\operatorname{tn}(x)-1}], [y_1 \ldots y_{\operatorname{tn}(x)-1}]) = 2(\operatorname{tn}(x) - 1).$$

On the other hand, for all $i > \operatorname{tn}(x)$ either have $z_i = x_i$ or $z_i = 1 - x_i = y_i$. Thus, the last last $n - \operatorname{tn}(x)$ bits of $x$ and $y$ contribute exactly $n - \operatorname{tn}(x)$ to the sum $\operatorname{eq}(z,x) + \operatorname{eq}(z,y)$; formally,

$$\operatorname{eq}([z_{\operatorname{tn}(x)+1} \ldots z_n], [x_{\operatorname{tn}(x)+1} \ldots x_n]) + \operatorname{eq}([z_{\operatorname{tn}(x)+1} \ldots z_n], [y_{\operatorname{tn}(x)+1} \ldots y_n]) = n - \operatorname{tn}(x).$$

By definition we also have $y_{\operatorname{tn}(x)} = x_{\operatorname{tn}(x)}$ and, thus,

$$\begin{aligned}
&\operatorname{eq}(z,x) + \operatorname{eq}(z,y) \\
&= \operatorname{eq}([z_1 \ldots z_{\operatorname{tn}(x)-1}], [x_1 \ldots x_{\operatorname{tn}(x)-1}]) + \operatorname{eq}(z_{\operatorname{tn}(x)}, x_{\operatorname{tn}(x)}) + \operatorname{eq}([z_{\operatorname{tn}(x)+1} \ldots z_n], [x_{\operatorname{tn}(x)+1} \ldots x_n]) \\
&\quad + \operatorname{eq}([z_1 \ldots z_{\operatorname{tn}(x)-1}], [y_1 \ldots y_{\operatorname{tn}(x)-1}]) + \operatorname{eq}(z_{\operatorname{tn}(x)}, x_{\operatorname{tn}(x)}) + \operatorname{eq}([z_{\operatorname{tn}(x)+1} \ldots z_n], [y_{\operatorname{tn}(x)+1} \ldots y_n]) \\
&= 2(\operatorname{tn}(x) - 1) + n - \operatorname{tn}(x) + 2\operatorname{eq}(z_{\operatorname{tn}(x)}, x_{\operatorname{tn}(x)}) \\
&= n + \operatorname{tn}(x) + 2\operatorname{eq}(z_{\operatorname{tn}(x)}, x_{\operatorname{tn}(x)}) - 2
\end{aligned}$$

This shows that $\operatorname{eq}(z,x) + \operatorname{eq}(z,y) = n + \operatorname{tn}(x)$ if and only if $\operatorname{eq}(z_{\operatorname{tn}(x)}, x_{\operatorname{tn}(x)}) = 1$, i.e., if and only if $z_{\operatorname{tn}(x)} = x_{\operatorname{tn}(x)} (= y_{\operatorname{tn}(x)})$.

It is immediate that for a secret code $z$ taken from $\{0,1\}^n$ uniformly at random, the probability to obtain, in one call of `LinAlg`, a string $y$ with $\mathrm{tn}(y) = \mathrm{tn}(x) + 1$ and $y_i = z_i$ for all $i < \mathrm{tn}(y)$ is $1/2$. This shows that, if Carole does not play a devil's strategy and if her string is taken from $\{0,1\}^n$ uniformly at random, we need, on average, two successive calls to procedure `LinAlg` until we obtain a string $y$ as desired.

Proposition 5 follows from the easy observation that it takes, on average, three iterations until both $y = x \oplus e^n_{\mathrm{tn}(x)}$, and $y = x \oplus \sum_{i=\mathrm{tn}(x)+1}^{n} e^n_i$ have been sampled. That is, even if Carole plays a devil's strategy, three calls of Algorithm 5 force her to accept one entry $z_{\mathrm{tn}(x)} \in \{0,1\}$. □

### 4.1.2 The case of $k \geq 3$ colors $\mathcal{C} = [0..k-1]$

The main argument of Proposition 5, namely that $\sum_{c=0}^{k-1} \mathrm{eq}(z, [c\ldots c]) = n$, seems hard to extend to more than two colors with no additional memory. However, having more than two colors can be exploited in a different way as it gives more than one way to mark the *tail* $[x_{\mathrm{tn}(x)} \ldots x_n]$ of a search point $x$.

**Proposition 6.** *For $k \geq 3$ colors, Algorithm 6 satisfies the claims of Lemma 4.*

---

**Algorithm 6:** Routine `LinAlg` for $k \geq 3$ colors

1 **Assumption:** The string $x \in \mathcal{C}^n$ in the memory satisfies $\mathrm{tn}(x) < n$ and $x_i = z_i$ for all $i < \mathrm{tn}(x)$;
2 With probability $(k-1)/k$ sample $y \in \{[x_1 \ldots x_{\mathrm{tn}(x)-1}|j|x_{\mathrm{tn}(x)+1} \ldots x_n] \mid j \in \mathcal{C}\setminus\{x_{\mathrm{tn}(x)}\}\}$ uniformly at random and with probability $1/k$ sample
$y \in \{[x_1 \ldots x_{\mathrm{tn}(x)-1}|j \ldots j] \mid j \in \mathcal{C}\setminus\{x_{\mathrm{tn}(x)-1}\}\}$ uniformly at random;
3 Query $\mathrm{eq}(z,y)$;
4 **if** $y = [x_1 \ldots x_{\mathrm{tn}(x)-1}|j|x_{\mathrm{tn}(x)+1} \ldots x_n]$, $j \neq x_{\mathrm{tn}(x)}$ **then**
5 $\quad$ **if** $\mathrm{eq}(z,y) > \mathrm{eq}(z,x)$ **then** $\mathcal{M} \leftarrow \{(y, \mathrm{eq}(z,y))\}$; $//z_{\mathrm{tn}(x)} = j$
6 **else**
7 $\quad \mathcal{M} \leftarrow \{(y, \mathrm{eq}(z,y))\}$;

---

*Proof.* Let $x \in \mathcal{C}^n$ with $\mathrm{tn}(x) < n$ and $x_i = z_i$ for all $i < \mathrm{tn}(x)$. If $y = [x_1 \ldots x_{\mathrm{tn}(x)-1}|j|x_{\mathrm{tn}(x)+1} \ldots x_n]$, then clearly we have $\mathrm{eq}(z,y) > \mathrm{eq}(z,x)$ if and only if $y_{\mathrm{tn}(x)} = j = z_{\mathrm{tn}(x)}$. Therefore, all we need to show is that, using the strategy of Algorithm 6, it takes a constant number of guesses until for each $j \in \mathcal{C}$ there exists an $i_j \in \mathcal{C}\setminus\{j\}$ such that we have queried both $x = [z_1 \ldots z_{\mathrm{tn}(x)-1}|i_j \ldots i_j]$ and $y = [z_1 \ldots z_{\mathrm{tn}(x)-1}|j|i_j \ldots i_j]$ in two subsequent guesses. This follows essentially from the fact that $k$ is constant.

More precisely—regardless of the current search point $x$—for any bitstring $y = [z_1 \ldots z_{\mathrm{tn}(x)-1}|j|i_j \ldots i_j]$ the probability to sample $y$ in the second of two subsequent calls to Algorithm 6 is constant. Therefore, the expected number of calls to Algorithm 6 until $y$ is sampled is constant. The claim follows by the linearity of expectation. □

### 4.2 Proof of Theorem 1

Building on `LinAlg` and the block-wise random guessing strategy introduced in Section 3, we can now present Paul's winning strategy for the single memory cell setting, which proves Theorem 1.

*Proof of Theorem 1.* The structure of this proof is as follows. First we sketch the main ideas and give a high-level pseudo-code for the size-one memory-restricted strategy winning the black answer-peg only Mastermind game with $k$ colors in $O(n/\log n)$ guesses. After fixing some notation, we then present more details for the different phases, in particular for the random guessing phase, which is the most critical part of this proof. We present here the details of Paul's strategy for the case of $k = 2$ colors. The generalization to $k \geq 3$ colors is pretty much straightforward. Some remarks on the differences between the case of $k = 2$ and $k \geq 3$ colors are given at the end of this proof.

Let us begin with the rough overview over Paul's strategy. He determines the first $n - \Theta(n/\log n)$ positions using random guessing, where he manages to store the random substrings and Carole's answers in the yet undetermined part of his one string in the memory. As in the proof of Theorem 2, he does so by iteratively determining blocks of length $s := \lceil \sqrt{n} \rceil$. Then, using the linear time strategy from Lemma 4, he determines the missing $\Theta(n/\log n)$ entries in $O(n/\log n)$ guesses.

To distinguish between the sampling and the linear time phase, Paul uses the last two entries $\texttt{suffix}(x) := [x_{n-1}x_n]$ of his string $x$. He has $\texttt{suffix}(x) = [01]$, when he is in the random guessing phase, and he uses $\texttt{suffix}(x) = [cc]$ for some $c \in \mathcal{C}$ to indicate that he applies calls to $\texttt{LinAlg}$. Once Paul has determined all but the last two entries (visible from $\text{tn}(x) = n - 1$), he simply needs to sample uniformly at random from the set of all $k^2 - 1$ remaining possible strings. This clearly determines $z$ in a constant expected number of additional queries (phase 3).

The total expected number of guesses can be bounded by

$$\underbrace{\frac{n-2}{s}(1 - \Theta(\log^{-1} n))}_{\substack{\text{number of blocks de-}\\\text{termined in phase 1}}} \underbrace{O(\tfrac{s}{\log s})}_{\substack{\text{queries needed to de-}\\\text{termine any such block}}} + \underbrace{O(\tfrac{n}{\log n})}_{\substack{\text{queries needed}\\\text{in phase 2}}} + \underbrace{O(1)}_{\substack{\text{queries needed}\\\text{in phase 3}}} = O(\tfrac{n}{\log n}).$$

A non-trivial part is the random guessing phase. As in the proof of Theorem 2, after guessing $t + k$ strings, we want to be able to regain the full guessing history. If we simply stored the random substring and Carole's reply in some unused part of $x$, then this changed memory would influence Carole's next answer and we would be unable to deduce information on the next guess from it. We solve this difficulty as follows. We store Carole's latest reply (i.e., value $\text{eq}(z, x)$ currently in the memory) and we sample new (random) substrings for the current block at the same time. Here we store the value $\text{eq}(z, x)$ in a part of $x$, for which we know the entries of Carole's hidden code. By this, we can separate in Carole's next answer the influence of the just stored information from the one of the random guess. The precise description of this $\texttt{Sampling}$ strategy is presented below.

To gain the storage space, for which we know the hidden code, we need to add another phase, phase 0, in which we apply $O(\log n)$ calls to the $\texttt{LinAlg}$ procedure until we have determined the first $\ell := \ell_n + 1$ positions of $z$ (cf. Lemma 4). This does not change the overall asymptotic number of queries Paul needs to win the game.

The pseudo-code for this size-one memory-restricted strategy is given in Algorithm 7. Similar to the notation in the proof of Theorem 2, we denote for any $h \in [0..n]$ its binary encoding of length $\ell_n$ by $\texttt{binary}_{\ell_n}(h)$ and for $h \in [0..s]$ we denote its binary encoding of length $\ell_s := \lceil \log s \rceil + 1$ by $\texttt{binary}_{\ell_s}(h)$. The current block of interest $i(x)$ is encoded in positions $\{n - \ell_s - 1, \ldots, n - 2\}$, i.e., we have $i(x) := \sum_{h=0}^{\ell_s - 1} 2^h x_{n-2-h}$, $B_{i(x)} := \{\ell + (i(x) - 1)s + 1, \ldots, \ell + i(x)s\}$, and, consequently, $\texttt{BLOCK}_{i(x)}(x) := [x_{\ell + (i(x)-1)s+1} \ldots x_{\ell + i(x)s}]$. The total number of blocks, which we determine via random guessing, is $b := \lfloor \frac{n-2}{s}(1 - \frac{K}{\log n}) \rfloor$ for some suitable large constant $K$. The number of random guesses for each block is $t := \lceil (2 + \varepsilon) \frac{s(1+2\log k)}{\log s - \log k} \rceil$ where $\varepsilon > 0$ is an arbitrarily small constant. Lastly, the actual number of already sampled guesses for block $B_{i(x)}$ is denoted by $q(x)$. As in the proof of Theorem 2, $q(x)$ can be computed via $p_1(x) := \max\{i \in [n - \ell_s - 3] \mid x_i = 1\}$, the largest position $i < n - 2 - \ell_s$ with entry $x_i = 1$. Details how $q(x)$ can be computed are given in the description of the OptimizeBlock routine, which, after $t$ random samples have been sampled via the Sampling routine, determines $\text{BLOCK}_{i(x)}(z)$, stores it in $B_{i(x)}$ and increases the block counter $i(x)$ by one.

---

**Algorithm 7:** A size-one memory-restricted algorithm winning the $k$-color black answer-peg only Mastermind game in $O(n/\log n)$ guesses.

---

**1** **Initialization:** Let $\mathcal{M} \leftarrow \emptyset$;
**2** **if** $\mathcal{M} = \emptyset$ **then**
**3** $\quad$ $x \leftarrow [c \ldots c]$ for some $c \in \mathcal{C}$ chosen u.a.r.;
**4** $\quad$ Query $\text{eq}(z, x)$ and update $\mathcal{M} \leftarrow \{(x, \text{eq}(z, x))\}$;
**5** **if** $\exists c \in \mathcal{C} : \textit{suffix}(x) = [cc] \wedge \text{tn}(x) \leq \ell$ **then**
**6** $\quad$ LinAlg; //find the first $\ell$ entries $[z_1 \ldots z_\ell]$
**7** **else if** $\exists c \in \mathcal{C} : \textit{suffix}(x) = [cc] \wedge \text{tn}(x) = \ell + 1$ **then**
**8** $\quad$ $x \leftarrow [\underbrace{0 \ldots 0}_{\ell} | \underbrace{0 \ldots 0}_{bs} | x_1 \ldots x_\ell | \underbrace{0 \quad \ldots \quad 0}_{n-(2\ell+bs+\ell_s+2)} | \text{binary}_{\ell_s}(1) | 01]$; //copy prefix (which coincides with the hidden code)
**9** $\quad$ Query $\text{eq}(z, x)$ and update $\mathcal{M}$ by replacing $(x, \text{eq}(z, x))$;
**10** **else if** $\textit{suffix}(x) = [01] \wedge i(x) \leq b \wedge q(x) < t + k$ **then**
**11** $\quad$ Apply Sampling;
**12** **else if** $\textit{suffix}(x) = [01] \wedge i(x) \leq b \wedge q(x) = t + k$ **then**
**13** $\quad$ Apply OptimizeBlock;
**14** **else if** $\textit{suffix}(x) = [01] \wedge i(x) = b + 1$ **then**
**15** $\quad$ $x \leftarrow [x_{\ell+bs+s+1} \ldots x_{2\ell+bs+s+1} | x_{\ell+1} \ldots x_{\ell+bs} | c \ldots c]$ with $c \in \mathcal{C} \setminus \{x_{\ell+bs}\}$ u.a.r.;
**16** $\quad$ Query $\text{eq}(z, x)$ and update $\mathcal{M}$ by replacing $(x, \text{eq}(z, x))$; //prepares $x$ for LinAlg
**17** **else if** $\exists c \in \mathcal{C} : \textit{suffix}(x) = [cc] \wedge \ell + bs < \text{tn}(x) \leq n - 2$ **then**
**18** $\quad$ LinAlg;
**19** **else if** $\exists c \in \mathcal{C} : \textit{suffix}(x) = [cc] \wedge \text{tn}(x) = n - 1$ **then**
**20** $\quad$ Sample $y \in \{[x_1 \ldots x_{n-2}|p] \mid p \in \mathcal{C}^2\} \setminus \{x\}$ uniformly at random;
**21** $\quad$ Query $\text{eq}(z, y)$;
**22** $\quad$ **if** $\text{eq}(z, y) = n$ **then** $\mathcal{M} \leftarrow \{(y, \text{eq}(z, y))\}$; //secret code found
**23** Go to line 2;

---

Let us now present a more detailed description of Algorithm 7.

As in the proof of Theorem 2 let us assume that Carole has chosen a fix code $z \in \mathcal{C}^n$, which she does not change during the game, i.e., to any of Paul's guesses $x$ she replies $\text{eq}(z, x)$. By adopting a worst-case view below, we implicitly still allow Carole to change $z$ as long as the new choice is consistent with all previous replies.

If in any iteration we find an $x$ with $\text{eq}(z, x) = n$, we have $x = z$ and all we need to do is to output $x$. Thus, in what follows we always assume $\text{eq}(z, x) < n$.

**Initialization of Algorithm 7, lines 1–4.** For initialization, Paul picks a $c \in \mathcal{C}$ uniformly at random and guesses the all-"c"s string of length $n$, $x = [c \ldots c]$. He updates the memory $\mathcal{M} \leftarrow \{(x, \text{eq}(z, x))\}$ accordingly. This memory satisfies all conditions of line 1 of routine `LinAlg` (Algorithms 5 and 6) with $\text{tn}(x) = 1$.

**Phase 0 of Algorithm 7, lines 5–6.** To this string, Paul applies successive calls to the routine `LinAlg`. By Lemma 4 he finds a string $y \in \mathcal{C}^n$ with $y_i = z_i$, $i \leq \ell$, and $\text{tn}(y) = \ell + 1$ in an expected number of $O(\ell)$ guesses. As mentioned above, Paul runs this first phase until he has determined the first

$$\ell = \lceil \log n \rceil + 2 = \ell_n + 1$$

entries. This is the number

$$\ell_n = \lceil \log n \rceil + 1$$

of entries needed to store in binary any integer value $h \in [0..n]$ plus 1. These bits shall be used in phase 1 of Algorithm 7 to indicate the status of the `Sampling` routine (first position) and for storing Carole's latest reply $\text{eq}(z, x) \in [0..n]$ (positions $\{2, \ldots, \ell\}$). We shall describe this in more detail below.

**Intermediate step, lines 7–9.** After Paul has determined the first $\ell$ entries, he needs to prepare the string for the random guessing phase, which is the main part of Algorithm 7. Since we want to use the first $\ell$ entries to store reference values, we need to make a copy of the prefix (which, by construction, coincides with Carole's hidden code). To this end, we query in line 7 the string

$$y = [\ \underbrace{0 \ldots 0}_{\ell + bs \text{ entries}}\ |x_1 \ldots x_\ell|\ \underbrace{0 \quad \ldots \quad 0}_{n - (2\ell + bs + \ell_s + 2) \text{ entries}}\ |\underbrace{\texttt{binary}_{\ell_s}(1)}_{\ell_s \text{ entries}}|01\ ],$$

where $x$ is the string that is currently in the memory, i.e., the string we obtained through phase $0^1$ and $bs = n - \Theta(n/\log n)$ is the number of positions Paul determines using random guessing. As mentioned in the overview, the last two entries $\texttt{suffix}(x) = [x_{n-1} x_n] = [01]$ indicate that we are entering the second phase. Note that throughout the game we have

$$\texttt{suffix}(x) = \begin{cases} [01], & \text{if we are in the second phase of the algorithm} \\ [cc], & \text{for some } c \in \mathcal{C}, \text{ otherwise.} \end{cases}$$

In positions $\{n - \ell_s - 1, \ldots, n - 2\}$ we indicate in binary the block, which we are currently trying to determine. That is, whenever $x$ is the current string in the memory with $\texttt{suffix}(x) = [01]$, then the block currently of interest is $B_{i(x)}$ with $i(x) = \sum_{i=0}^{\ell_s - 1} 2^i x_{n-2-i}$. Here in this intermediate step we initialize $i(x) = 1$.

After guessing $y$ and updating the memory by replacing the current one with $\{(y, \text{eq}(z, y))\}$, Paul enters the first phase (as indicated by $\texttt{suffix}(x)$). The overall expected number of queries needed until this point is $O(\ell) = O(\log n)$.

**First phase of Algorithm 7, lines 10–13.** The first phase is the main phase of Algorithm 7. In this phase, Paul determines all but $n - \Theta(n/\log n)$ entries by iteratively determining blocks of length $s = \lceil \sqrt{n} \rceil$ via random guessing. In total, he determines $b = \lfloor \frac{n}{s}(1 - \frac{K}{\log n}) \rfloor$ such blocks in this phase. The description of the routines `Sampling` (in which $k$ reference strings and $t$ random samples are queried for the $i(x)$-th block $B_{i(x)}$) and `OptimizeBlock` (in which we use the reference strings and the random guesses to determine $\text{BLOCK}_{i(x)}(z)$, the $i(x)$-th block of the secret code $z$) is quite technical. We present the details after the description of the remaining phases.

---

[1]That is, we have $\text{tn}(x) = \ell + 1$, $\forall i \leq \ell : x_i = z_i$, and $\exists c \in \mathcal{C} \setminus \{x_\ell\} \forall i \geq \text{tn}(x) : x_i = c$.

**Second phase of Algorithm 7, lines 14–18.** In the second phase of Algorithm 7 we again apply successive calls to routine LinAlg to determine all but the last two remaining entries. To this end, we first need to prepare the string. This is done in lines 14–16 of Algorithm 7. It follows from the correctness of the first phase that the string $x$ queried in line 16 satisfies $x_i = z_i$ for all $1 \leq i \leq \ell + bs$. And, by definition, it also satisfies $x_{\text{tn}(x)-1} \neq x_{\text{tn}(x)}$ with $\text{tn}(x) = \ell + bs + 1$.

From Lemma 4 we infer that via routine LinAlg we find a string $x$ with $\text{tn}(x) = n - 1$ and $x_i = z_i$ for all $1 \leq i \leq n - 2$ in an expected number of $O(n - 2 - (\ell + bs)) = O(n/\log n)$ queries. These are lines 17 and 18 of Algorithm 7.

**Third phase of Algorithm 7, lines 19–22.** Comparable to the last step of Algorithm 4, all we need to do in the last phase of Algorithm 7 is to determine the last two entries. This is done by sampling $y$ uniformly at random from the set of possible target strings $\{[x_1 \ldots x_{n-2}|p] \mid p \in \{0,1\}^2\} \setminus \{x\}$ and we find $y = z$ after a constant expected number of queries. This phase is recognized by the algorithm by the fact that $\text{tn}(x) = n - 1$. Note that we have $\text{tn}(x) \leq n - 2$ in the LinAlg phases—phases 0 and 2—and that we have $\text{tn}(x) = n$ in phase 1.

Summing up the expected number of queries needed for each phase, we have shown that Paul needs, on average,

$$O(\log n) + O(n/\log n) + O(n/\log n) + O(1) = O(n/\log n)$$

until he has identified Carole's hidden code $z$.

In the remainder of this proof we present the details of the first phase of Algorithm 7, the random sampling routine Sampling and the OptimizeBlock routine. As mentioned above, this description requires some technicalities. Therefore, we split it into the following parts:

> **Part I** In the first part, we present the general structure of the guesses Paul queries in the sampling phase. Here, we shall also show that the $n$ positions are indeed sufficient to store, for any of the $b$ blocks of length $s$ all necessary information about the samples.
>
> **Part II** The second part, which is brief, provides further notation used in the pseudo-code of Algorithm 8.
>
> **Part III** The main part is the third one. Here we show how the contributions $\Delta_{i(x)}(r) \in [0..s]$ of the random samples $r \in \mathcal{C}^s$ can be computed. This also shows that indeed after sampling the $t$ random guesses for the current block of interest, it is possible to regain the full query history using only the information that has been stored in the memory. This is clearly the most technical part of this proof.
>
> **Part IV** We conclude the description of phase 1 in the fourth part, where we explain how the memory is being updated, once the entries $\text{BLOCK}_{i(x)}(z)$ of the secret code $z$ in the $i(x)$-th block have been determined.

**Part I** The general structure of a random query $x$ for determining block $B_{i(x)}$ is the following.

$$x = [\underbrace{x_1}_{(1)} \,|\, \underbrace{\mathtt{binary}_{\ell_n}(\mathrm{eq}(z,y))}_{(2)} \,|\, \underbrace{\mathtt{opt}(B_1)|\ldots|\mathtt{opt}(B_{i(x)})}_{(3)} \,|\, \underbrace{r}_{(4)} \,|\, \underbrace{0\ldots 0}_{(5)} \,|\, \underbrace{z_1\ldots z_\ell}_{(6)} \,| \quad (1)$$

$$\underbrace{\mathtt{binary}_{\ell_n}(\mathrm{eq}(z,x^0))|\mathtt{binary}_{\ell_n}(\mathrm{eq}(z,x^1))|1\,|}_{(7)}$$

$$\underbrace{\mathtt{binary}_{\ell_n}(\mathrm{eq}(z,\mathtt{ref}^{(1)}))|r^{(1)}|\Delta_{i(x)}(r^{(1)})|1|\ldots|}_{(8)}$$

$$\underbrace{\mathtt{binary}_{\ell_n}(\mathrm{eq}(z,\mathtt{ref}^{(t')}))|r^{(t')}|\Delta_{i(x)}(r^{t'})|1\,|0\ldots 0|}_{(8)\text{ continued}} \underbrace{\mathtt{binary}_{\ell_s}(i(x))}_{(9)} \,|\, \underbrace{01}_{(10)}\,],$$

where we use

(1) the first entry $x_1 \in \{0,1\}$ to indicate whether we are sampling a new random substring ($x_1 = 1$) or whether we are doing a storage operation only through which we add to $x$ all necessary information from the previous guess($x_1 = 0$). An explanation of these operation follows below;

(2) $\ell_n$ entries for encoding the value $\mathrm{eq}(z,y) \in [0..n]$ of the string $y$ that is currently stored in the memory cell (serves as reference value),

(3) $(i(x)-1)s$ entries for the already determined blocks $B_1, \ldots, B_{i(x)-1}$,

(4) $s$ entries for the current block $B_{i(x)}$ of interest. If we are sampling new information (i.e., if $x_1 = 1$), then the substring $r$ is a string taken from $\mathcal{C}^s$ uniformly at random and $r$ is the all-zeros string of length $s$ otherwise;

(5) $(b-i(x))s$ zeros for the yet untouched blocks $B_{b'}$ with $i(x) < b' \leq b$,

(6) $\ell$ entries for storing the length-$\ell$ prefix that coincides with Carole's hidden code (obtained trough phase 0),

(7) $2\ell_n + 1$ entries for storing the values $\mathrm{eq}(z,x^0)$ and $\mathrm{eq}(z,x^1)$ of the two reference strings $x^0$ and $x^1$ (explanation follows),

(8) $t'(\ell_n + s + \ell_s + 1)$, $t' \leq t$, entries for storing, for each random sample, (i) the value $\mathrm{eq}(z,\mathtt{ref})$ for a reference string $\mathtt{ref}$ (in binary, requires $\ell_n$ positions), (ii) the random sample $r \in \mathcal{C}^s$ itself, (iii) its contribution $\Delta_{i(x)}(r) \in [0..s]$ to Carole's reply (in binary, requires $\ell_s$ positions), and (iv) one additional "1" (to ease the computation of the number of guesses $q(x)$ via $p_1(x)$; details follow),

(9) $\ell_s$ entries for encoding in binary, which block we are currently trying to determine, and

(10) the last two entries, $\mathtt{suffix}(x)$, for indicating the current phase of the algorithm.

Clearly, one critical part is the limited storage capacity. For this reason, let us show that we have enough positions to store all the information needed to compute $\mathcal{S}_{i(x)}^{\text{consistent}}$, the set of all strings consistent with Carole's replies for the random guesses in the $i(x)$-th block $B_{i(x)}$.

Recall that, by Theorem 3, for determining the $i(x)$-th block $\texttt{BLOCK}_{i(x)}(z)$ of $z$, we need $t = (2+\varepsilon)\frac{s(1+2\log k)}{\log s - \log k} = \Theta(s/\log n)$ random guesses ($\varepsilon > 0$ being an arbitrarily small constant). In addition, equivalently to the proof of Theorem 2, we need again 2 reference strings $x^0$ and $x^1$ (reference (7) in equation (1)). These two reference strings will be needed to infer the contributions $\Delta_{i(x)}(r)$ of the random samples $r \in \mathcal{C}^s$ in the $i(x)$-th block.

From the structure of the guesses presented in equation (1) above, we infer that the total storage requirement can be bounded from above by

$$1 + \ell_n + bs + \ell + 2\ell_n + 1 + t(\ell_n + s + \ell_s + 1) + \ell_s + 2$$
$$= bs + ts + o(n/\log n) \leq n\bigl(1 - \frac{K}{\log n}\bigr) + \Theta(n/\log n) + o(n/\log n) < n$$

for sufficiently large, but constant $K$ and sufficiently large $n$. This shows that, for sufficiently large $n$, Paul indeed can store all information needed to compute $\mathcal{S}_{i(x)}^{\text{consistent}}$ in one single string of length $n$.

**Part II** Let us now fix the notation used in the pseudo-code of routine $\texttt{Sampling}$ (Algorithm 8). For all $b' < i(x)$ we set

$$\texttt{opt}(B_{b'}) := [x_{\ell+(b'-1)s+1} \ldots x_{\ell+b's}],$$

the entries of $x$ in the $b'$-th block. The notation "$\texttt{opt}$" is justified by the fact that we shall have $\texttt{opt}(B_{b'}) = \texttt{BLOCK}_{b'}(z)$ for all $b' < i(x)$. Furthermore, let

$$\texttt{AddReferenceStringInfo}(x) := [\underbrace{0 \ldots 0}_{(1),(2)} | \underbrace{\texttt{opt}(B_1) | \ldots | \texttt{opt}(B_{i(x)-1})}_{(3)} | \underbrace{0 \ldots 0}_{(4)} |$$
$$\underbrace{x_{\ell+i(x)s+1} \ldots x_{2\ell+bs}}_{(5),(6)} | \underbrace{x_2 \ldots x_\ell | \texttt{binary}_{\ell_n}(\text{eq}(z,x)) | 1}_{(7)} | \underbrace{x_{2\ell+bs+2\ell_n+2} \ldots x_n}_{(*)}],$$

where the references in the expression above are the same as the ones used in equation (1) and where $(*)$ is simply a copy of the last entries of $x$. That is, the $\texttt{AddReferenceStringInfo}(x)$ operation adds to $x$ the values $\text{eq}(z, x^0)$ and $\text{eq}(z, x^1)$ to the memory and each of these values is stored in binary notation of length $\ell_n$. Lastly, we denote by $\texttt{Add}(\text{eq}(z,x))$ the operation

$$\texttt{Add}(\text{eq}(z,x)) := [\underbrace{0 \ldots 0}_{(1),(2)} | \underbrace{\texttt{opt}(B_1) | \ldots | \texttt{opt}(B_{i(x)-1})}_{(3)} | \underbrace{0 \ldots 0}_{(4)} | \underbrace{x_{\ell+i(x)s+1} \ldots x_{p_1(x)}}_{(5),(6),(7),(8)} | \quad (2)$$
$$\underbrace{x_2 \ldots x_\ell | \texttt{BLOCK}_{i(x)}(x) | \texttt{binary}_{\ell_s}(\Delta_{i(x)}(\texttt{BLOCK}_{i(x)}(x))) | 1}_{(\dagger)} | \underbrace{x_{p_1(x)+\ell_n+s+\ell_s+2} \ldots x_n}_{(*)}],$$

which adds (in substring $(\dagger)$) to the memory

- a copy of the value $\text{eq}(z, \texttt{ref})$ of a reference string $\texttt{ref}$ (which was previously stored in positions $\{2, \ldots, \ell\}$),
- the random sample $\texttt{BLOCK}_{i(x)}(x)$ of the last guess,
- the contribution $\Delta_{i(x)}(\texttt{BLOCK}_{i(x)}(x))$ of the random sample $\texttt{BLOCK}_{i(x)}(x)$ to $\text{eq}(z, x)$, and
- the one additional "1" that shall ease the computation of $q(x)$, the number of already queried samples.

All other but the first $\ell$ entries (which are set to zero) are copied from $x$. The references in equation (2) are the same as in equation (1).

---

**Algorithm 8:** The Sampling routine for $k = 2$ colors.

1 **Assumption:** Memory $\mathcal{M} = \{(x, \text{eq}(z, x))\}$ satisfies $\text{eq}(z, x) < n$, $\texttt{suffix}(x) = [01]$, $i(x) \leq b$, and $q(x) < t + 2$;
2 **if** $q(x) = 0 \wedge x_1 = 0$ **then**
3 $\quad x \leftarrow [1|\texttt{binary}_{\ell_n}(\text{eq}(z, x))|\texttt{opt}(B_1)|\ldots|\texttt{opt}(B_{i(x)-1})|1\ldots 1|x_{\ell+i(x)s+1}\ldots x_n]$;
4 $\quad$ Query $\text{eq}(z, x)$ and update $\mathcal{M}$ by replacing $(x, \text{eq}(z, x))$;
5 **else if** $q(x) = 0 \wedge x_1 = 1$ **then**
6 $\quad x \leftarrow \texttt{AddReferenceStringInfo}(x)$ ;
7 $\quad$ Query $\text{eq}(z, x)$ and update $\mathcal{M}$ by replacing $(x, \text{eq}(z, x))$;
8 **else if** $2 \leq q(x) < t + 2 \wedge x_1 = 0$ **then**
9 $\quad x \leftarrow [1|\texttt{binary}_{\ell_n}(\text{eq}(z, x))|\texttt{opt}(B_1)|\ldots|\texttt{opt}(B_{i(x)-1})|r|x_{\ell+i(x)s+1}\ldots x_n]$ for $r \in \mathcal{C}^s$ chosen u.a.r.;
10 $\quad$ Query $\text{eq}(z, x)$ and update $\mathcal{M}$ by replacing $(x, \text{eq}(z, x))$;
11 **else if** $2 \leq q(x) < t + 2 \wedge x_1 = 1$ **then**
12 $\quad x \leftarrow \texttt{Add}(\text{eq}(z, x))$;
13 $\quad$ Query $\text{eq}(z, x)$ and update $\mathcal{M}$ by replacing $(x, \text{eq}(z, x))$;

---

**Part III** Let us now show in detail how to infer the contributions $\Delta_{i(x)}(r)$ of the random guesses. For clarity, we show how to do this for the first block, i.e., for the positions $\{\ell+1\ldots\ell+s\}$. The procedure is similar for all other blocks and we shall comment on this case at the end of this part.

First note that after the intermediate step in lines 7 to 9 of Algorithm 7, Paul enters the routine Sampling with $\mathcal{M} = \{(x^0, \text{eq}(z, x^0))\}$, where

$$x^0 = [\underbrace{0 \ldots 0}_{\ell+bs \text{ entries}}|z_1 \ldots z_\ell| \underbrace{0 \ldots 0}_{n-(2\ell+bs+\ell_s+2) \text{ entries}} |\underbrace{\texttt{binary}_{\ell_s}(1)}_{\ell_s \text{ entries}}|01],$$

and he queries in the first sampling iteration (lines 2–4 of Algorithm 8)

$$x^1 = [\underbrace{1|\texttt{binary}_{\ell_n}(\text{eq}(z, x^0))}_{1+\ell_n=\ell \text{ entries}}|\underbrace{1\ldots 1}_{s \text{ entries}}|\underbrace{0 \ldots 0}_{(b-1)s \text{ entries}}|z_1 \ldots z_\ell| \underbrace{0 \ldots 0}_{n-(2\ell+bs+\ell_s+2) \text{ entries}}|\texttt{binary}_{\ell_s}(1)|01]$$

with the all-ones substring in the first block. We can compute the contribution of the first $\ell$ entries $[1|\texttt{binary}_{\ell_n}(\text{eq}(z, x^0))]$ to the value $\text{eq}(z, x^1)$ via

$$\tilde{f}(x^1) := \text{eq}([z_1 \ldots z_\ell], [1|\texttt{binary}_{\ell_n}(\text{eq}(z, x^0))]) = \text{eq}([x^1_{\ell+bs+1} \ldots x^1_{2\ell+bs}], [x^1_1 \ldots x^1_\ell]),$$

and, by the same reasoning, the contribution of the first $\ell$ entries in $x^0$ to $\text{eq}(z, x^0)$ via $\tilde{f}(x^0) = \text{eq}([x^0_{\ell+bs+1} \ldots x^0_{2\ell+bs}], [0 \ldots 0])$. Let us, for a moment, assume that we now had $\mathcal{M} = \{(x^1, \text{eq}(z, x^1))\}$ and that we had another string

$$y = [z_1 \ldots z_\ell|\underbrace{r}_{s \text{ entries}}|\underbrace{0\ldots 0}_{(b-1)s \text{ entries}}|z_1 \ldots z_\ell|\underbrace{0\ldots 0}_{n-(2\ell+bs+\ell_s+2) \text{ entries}}|\texttt{binary}_{\ell_s}(1)|01]$$

for some random substring $r \in \mathcal{C}^s$. Then we could compute the contribution of the random entries $r$ in the first block $B_1$ of $y$ via

$$\tilde{\Delta}_1(r) = \text{eq}(z,y) - \frac{\text{eq}(z,x^0) + \text{eq}(z,x^1) + (\ell - \tilde{f}(x^0)) + (\ell - \tilde{f}(x^1)) - s}{2}.$$

Key to this equality is the fact that the first $\ell$ entries of $y$ contribute $\ell$ to Carole's response $\text{eq}(z,y)$ to guess $y$, whereas the first $\ell$ entries of $x^0$ and $x^1$ contribute $\tilde{f}(x^0) + \tilde{f}(x^1)$ to the sum $\text{eq}(z,x^0) + \text{eq}(z,x^1)$ and the fact that the entries in the first block, $[0\ldots 0]$ and $[1\ldots 1]$, respectively, contribute in total $s$ towards $\text{eq}(z,x^0) + \text{eq}(z,x^1)$. All other entries $x_i, y_i, i > \ell + s$ contribute either 2 or 0 to the sum $\text{eq}(z,x^0) + \text{eq}(z,x^1)$ and every entry contributes 2 if and only if it contributes 1 to the value $\text{eq}(z,y)$.

Note however, that we would now have to choose now which of the strings to keep in the memory and we would eventually loose the information $\text{eq}(z,x^1)$. Therefore, in lines 6 and 7 in Algorithm 8, we first query the reference string

$$x^2 = [\,\underbrace{0\ldots 0}_{\ell \text{ entries}} \,|\, \underbrace{0\ldots 0}_{s \text{ entries}} \,|\, \underbrace{0\ldots 0}_{(b-1)s} \,|\, z_1 \ldots z_\ell |\, \underbrace{\texttt{binary}_{\ell_n}(\text{eq}(z,x^0))|\texttt{binary}_{\ell_n}(\text{eq}(z,x^1))|1}_{2\ell_n + 1 \text{ entries}} \,|$$

$$\underbrace{0\ldots 0}_{\substack{n - (2\ell + bs + 2\ell_n + \\ 1 + \ell_s + 2) \text{ entries}}} \quad |\texttt{binary}_{\ell_s}(1)|01\,].$$

This query is needed only to store the values $\text{eq}(z,x^0)$ and $\text{eq}(z,x^1)$ of both reference strings. Since adding the substring $[\texttt{binary}_{\ell_n}(\text{eq}(z,x^0))|\texttt{binary}_{\ell_n}(\text{eq}(z,x^1))|1]$ to the memory string again changes the number of positions, in which the guess and Carole's hidden string coincide, we need to store this information in the next query as well. More precisely, we have that $x^0$ and $x^2$ differ in exactly the substring $[\texttt{binary}_{\ell_n}(\text{eq}(z,x^0))|\texttt{binary}_{\ell_n}(\text{eq}(z,x^1))|1]$, and the contribution of this substring (compared to the all-zeros substring which it replaces) is $\text{eq}(z,x^2) - \text{eq}(z,x^0)$.

Furthermore, we need to indicate that we are sampling a new random substring. This is the first position in the string and the next $\ell - 1$ entries are needed to encode in binary the value $\text{eq}(z,x^2)$. That is, instead of querying $y$ as above we query (lines 9 and 10 in Algorithm 8)

$$x^3 = [1|\texttt{binary}_{\ell_n}(\text{eq}(z,x^2))|r^{(1)}|0\ldots 0|z_1\ldots z_\ell|\texttt{binary}_{\ell_n}(\text{eq}(z,x^0))|\texttt{binary}_{\ell_n}(\text{eq}(z,x^1))|1|$$
$$0\ldots 0|\texttt{binary}_{\ell_s}(1)|01],$$

where the substring $r^{(1)} \in \mathcal{C}^s$ in $B_1$ is taken uniformly at random. The number of zeros in the first all-zeros substring is again $(b-1)s$ and in the second all-zeros substring it is $n - (2\ell + bs + 2\ell_n + 1 + \ell_s + 2)$. Now, in the same fashion as above, we can compute the contribution $\Delta_1(r^{(1)}) = \text{eq}([z_{\ell+1}\ldots z_{\ell+s}])$ of the substring $r^{(1)} \in \mathcal{C}^s$ via

$$\Delta_1(r^{(1)}) = \text{eq}(z,x^3) - \Big(\frac{\text{eq}(z,x^0) + \text{eq}(z,x^1) + (\ell - \tilde{f}(x^1)) + (\ell - \tilde{f}(x^0)) - s}{2}$$
$$+ (\text{eq}(z,x^2) - \text{eq}(z,x^0)) - (\ell - \tilde{f}(x^3))\Big). \quad (3)$$

Also note that all the information needed for this computation is contained in the string $x^3$ itself.

Since later we want to be able to regain the full guessing history, in the next guess we store both the reference value $\text{eq}(z,x^2)$ as well as the contribution $\Delta_1(r^{(1)})$. And, of course, we also

**Algorithm 9:** The `OptimizeBlock` routine

1 **Assumption:** Memory $\mathcal{M} = \{(x, \text{eq}(z,x))\}$ satisfies $\text{eq}(z,x) < n$, $\text{suffix}(x) = [01]$, $i(x) \le b$, and $q(x) = t+2$;
2 **if** $x_1 = 0$ **then**
3     $y \leftarrow [1\ldots1|\text{opt}(B_1)|\ldots|\text{opt}(B_{i(x)-1})|w|x_{\ell+i(x)s+1}\ldots x_n]$ for $w \in \mathcal{S}_{i(x)}^{\text{consistent}}$ chosen u.a.r.;
4     Query $\text{eq}(z,y)$;
5     **if** $\Delta_{i(x)}(\text{BLOCK}_{i(x)}(y)) = s$ **then** $\mathcal{M} \leftarrow \{(y, \text{eq}(z,y))\}$; $//w = \text{BLOCK}_{i(x)}(z)$
6 **else**
7     $x \leftarrow \text{Update}(x)$;
8     Query $\text{eq}(z,x)$ and update $\mathcal{M}$ by replacing $(x, \text{eq}(z,x))$; //string prepared for determining the next block need to store the random guess $r^{(1)} = \text{BLOCK}_1(x^3)$ itself. Therefore, we query (lines 12 and 13 in Algorithm 8) in the next iteration of Algorithm 7

$$x^4 = [\underbrace{0\ldots0}_{\ell}|\underbrace{0\ldots0}_{s}|\underbrace{0\ldots0}_{(b-1)s}|z_1\ldots z_\ell|\text{binary}_{\ell_n}(\text{eq}(z,x^0))|\text{binary}_{\ell_n}(\text{eq}(z,x^1))|1|$$
$$\underbrace{\text{binary}_{\ell_n}(\text{eq}(z,x^2))}_{=[x_2^3\ldots x_\ell^3]}|\underbrace{\text{BLOCK}_1(x^3)}_{=r^{(1)}}|\underbrace{\text{binary}_{\ell_s}(\Delta_1(r^{(1)}))}_{\text{see equation (3)}}|1|0\ldots0|\text{binary}_{\ell_s}(1)|01].$$

Note that since $\Delta_1(r^{(1)}) \in [0..s]$, we can encode this value using $\ell_s$ positions only.

By continuing like this we are able to compute, in any iteration of the first phase, the contributions $\Delta_{i(x)}(r)$ of the random substrings $r \in \mathcal{C}^s$.

As in the proof of Theorem 2 we need to be able to compute how many random guesses have been queried already for the current block of interest. As indicated above, this can be derived from $p_1(x)$ as follows. For any random guess $r \in \mathcal{C}^s$ we use $\ell_n + s + \ell_s + 1$ entries for storing all information that will be needed later to regain the full guessing history. Furthermore, we used $2\ell_n + 1$ entries for storing the values $\text{eq}(z,x^0)$ and $\text{eq}(z,x^1)$ of the two reference strings $x^0$ and $x^1$, and we store information only in positions $i > 2\ell + bs$. Hence, the number of guesses for block $B_{i(x)}$ can be computed as

$$q(x) = \begin{cases} 0, & \text{if } p_1(x) \le 2\ell + bs \text{ and } x_1 = 0 \\ 1, & \text{if } p_1(x) \le 2\ell + bs \text{ and } x_1 = 1 \\ 2 + \frac{p_1(x)-(2\ell+bs+2\ell_n+1)}{\ell_n+s+\ell_s+1}, & \text{otherwise.} \end{cases}$$

After querying $t$ random guesses (i.e., after querying a total number of $t+k$ guesses) for the first block, we regain the full guessing history from the string $x$ currently in the memory as follows. The $i$-th random sample $r^{(i)} \in \mathcal{C}^s$, which we guessed for the first block is

$$r^{(i)} := [x_{2\ell+bs+2\ell_n+1+(i-1)(\ell_n+s+\ell_s+1)+\ell_n+1}\ldots x_{2\ell+bs+2\ell_n+1+(i-1)(\ell_n+s+\ell_s+1)+\ell_n+s}],$$

and the corresponding query was

$$y^{(i)} := [1|x_{2\ell+bs+2\ell_n+1+(i-1)(\ell_n+s+\ell_s+1)+1}\ldots x_{2\ell+bs+2\ell_n+1+(i-1)(\ell_n+s+\ell_s+1)+\ell_n}|$$
$$r^{(i)}|x_{\ell+s+1}\ldots x_{2\ell+bs+2\ell_n+1+(i-1)(\ell_n+s+\ell_s+1)}|0\ldots0|x_{n-\ell_s-1}\ldots x_n].$$

We have stored in binary the contribution $\Delta_1(r^{(i)})$ of $r^{(1)}$ to the overall function value $\text{eq}(z, y^{(i)})$ in positions

$$\{2\ell+bs+2\ell_n+1+(i-1)(\ell_n+s+\ell_s+1)+\ell_n+s+1 \ldots 2\ell+bs+2\ell_n+1+(i-1)(\ell_n+s+\ell_s+1)+\ell_n+s+\ell_s\}$$

and thus we have

$$\Delta_1(r^{(i)}) = \sum_{i=0}^{\ell_s-1} 2^i x_{2\ell+bs+2\ell_n+1+(i-1)(\ell_n+s+\ell_s+1)+\ell_n+s+\ell_s-i}.$$

By Theorem 3, the expected size of

$$\mathcal{S}_1^{\text{consistent}} := \{w \in \{0,1\}^s \mid \forall i \leq t : \text{eq}(y, r^{(i)}) = \Delta_1(r^{(i)})\}$$

is bounded from above by $1 + 1/s$. That is, we can now identify $\texttt{BLOCK}_1(z)$ in a constant number of guesses. These are lines 3 – 5 of routine $\texttt{OptimizeBlock}$ (Algorithm 9).

As mentioned above, determining the other blocks $2, \ldots, b$ is similar. In these iterations, the $(i(x)-1)s$ entries in positions $\{\ell+1, \ldots, \ell+(i(x)-1)s\}$ are already optimized, that is, they coincide with Carole's hidden string $z$. Thus, they are not changed in any further iteration of Algorithm 9.

**Part IV** Once $\texttt{BLOCK}_{i(x)}(z) = [z_{\ell+(i(x)-1)s+1} \ldots z_{\ell+i(x)s}]$ has been determined, we need to update the memory such that we can start determining the entries of the next block. These are lines 7 and 8 in Algorithm 9. Here we abbreviate

$$\texttt{Update}(x) := [\underbrace{0 \ldots 0}_{(a)} | \underbrace{\texttt{opt}(B_1) | \ldots | \texttt{opt}(B_{i(x)})}_{(b)} | \underbrace{0 \ldots 0}_{(c)} | \underbrace{x_{\ell+bs+1} \ldots x_{2\ell+bs}}_{(d)} |$$
$$\underbrace{0 \ldots 0}_{(e)} | \underbrace{\texttt{binary}_{\ell_s}(i(x)+1)}_{(f)} | \underbrace{01}_{(g)} ],$$

where

(a) the first $\ell$ entries are set to zero,

(b) we now have $i(x)$ already determined blocks,

(c) the new block of interest, block $i(x) + 1$, as well as all blocks $b' > i(x) + 1$ are (still) set to zero,

(d) we keep the copy of the prefix $[z_1 \ldots z_\ell]$ in positions $\{\ell + bs + 1, \ldots, 2\ell + bs\}$,

(e) all information that we have used in the previous query to determine block $B_{i(x)}$ is removed (and set to zero),

(f) the index for the current block of interest is increased by one, and

(g) the last two entries still indicate the second phase.

**The case of $k \geq 3$ colors.** For the general case, the main strategy as given by Algorithm 7 remains the same. What needs to be changed is the $\texttt{Sampling}$ routine, where instead of sampling only two reference strings $x^0$ and $x^1$, we need to sample $k$ reference strings $x^0, x^1, \ldots, x^{k-1}$ with $\texttt{BLOCK}_{i(x)}(x^c) = [c \ldots c]$ for all $c \in [0..k-1]$.

**Algorithm 10:** The Sampling routine for $k \geq 3$ colors.

**1 Assumption:** Memory $\mathcal{M} = \{(x, \text{eq}(z, x))\}$ satisfies $\text{eq}(z, x) < n$, $\text{suffix}(x) = [01]$, $i(x) \leq b$, and $q(x) < t + k$;

**2 if** $q(x) = 0 \wedge x_1 = 0$ **then**
**3** $\quad x \leftarrow [1|\text{binary}_{\ell_n}(\text{eq}(z, x))|\text{opt}(B_1)|\ldots|\text{opt}(B_{i(x)-1})|1\ldots 1|x_{\ell+i(x)s+1}\ldots x_n]$;
**4** $\quad$ Query $\text{eq}(z, x)$ and update $\mathcal{M}$ by replacing $(x, \text{eq}(z, x))$;

**5 else if** $q(x) = 0 \wedge x_1 = 1$ **then**
**6** $\quad x \leftarrow \text{AddReferenceStringInfo}(x)$ ;
**7** $\quad$ Query $\text{eq}(z, x)$ and update $\mathcal{M}$ by replacing $(x, \text{eq}(z, x))$;

**8 else if** $2 \leq q(x) < k \wedge x_1 = 0$ **then**
**9** $\quad x \leftarrow [1|\text{binary}_{\ell_n}(\text{eq}(z, x))|\text{opt}(B_1)|\ldots|\text{opt}(B_{i(x)-1})|q(x)\ldots q(x)|x_{\ell+i(x)s+1}\ldots x_n]$;
**10** $\quad$ Query $\text{eq}(z, x)$ and update $\mathcal{M}$ by replacing $(x, \text{eq}(z, x))$;

**11 else if** $2 \leq q(x) < k \wedge x_1 = 1$ **then**
**12** $\quad x \leftarrow \text{AddReferenceStringInfo}_2(x)$ ;
**13** $\quad$ Query $\text{eq}(z, x)$ and update $\mathcal{M}$ by replacing $(x, \text{eq}(z, x))$;

**14 else if** $k \leq q(x) < t + k \wedge x_1 = 0$ **then**
**15** $\quad x \leftarrow [1|\text{binary}_{\ell_n}(\text{eq}(z, x))|\text{opt}(B_1)|\ldots|\text{opt}(B_{i(x)-1})|r|x_{\ell+i(x)s+1}\ldots x_n]$ for $r \in \mathcal{C}^s$ chosen u.a.r.;
**16** $\quad$ Query $\text{eq}(z, x)$ and update $\mathcal{M}$ by replacing $(x, \text{eq}(z, x))$;

**17 else if** $k \leq q(x) < t + k \wedge x_1 = 1$ **then**
**18** $\quad x \leftarrow \text{Add}(\text{eq}(z, x))$;
**19** $\quad$ Query $\text{eq}(z, x)$ and update $\mathcal{M}$ by replacing $(x, \text{eq}(z, x))$;

Algorithm 10 shows the generalized sampling routine. Here we define

$$\text{AddReferenceStringInfo}_2(x) :=$$
$$[\underbrace{0\ldots 0}_{(1),(2)}|\underbrace{\text{opt}(B_1)|\ldots|\text{opt}(B_{i(x)-1})}_{(3)}|\underbrace{0\ldots 0}_{(4)}|\underbrace{x_{\ell+i(x)s+1}\ldots x_{p_1(x)}}_{(5),(6),(7),(7')}|$$
$$\underbrace{x_2\ldots x_\ell|\text{binary}_{\ell_n}(\text{eq}(z, x))|1}_{(\dagger)}|\underbrace{x_{p_1(x)+\ell_n+s+\ell_s+2}\ldots x_n}_{(*)}],$$

where

(1)–(7) are the same references as in equation (1),

(7') are the additional positions needed for storing the values $\text{eq}(z, x^j)$ of the already queried reference strings $x^2, \ldots, x^{q(x)-1}$ (each requiring $2\ell_n + 1$ positions),

(†) we add the information of the $q(x)$-th reference string $x^{q(x)}$ to the memory (again requiring $2\ell_n + 1$ positions), and

(∗) is simply a copy of the last entries of the previous guess.

The substring $[x_2 \ldots x_\ell]$ is needed again to infer the contribution of the positions, in which we added the information of the previous reference string $x^{q(x)-1}$. The reasoning is the same as in the case of $k = 2$ colors.

Since we added more reference string information, we need to adjust the definition of $q(x)$ accordingly. Since we need $2\ell_n + 1$ additional bits for each reference string $x^j$, $2 \leq j \leq k$, and we use $\ell_n + s + \ell_s + 1$ entries for storing the information of each random guess, we have

$$q(x) := \begin{cases} 0, & \text{if } p_1(x) \leq 2\ell + bs \text{ and } x_1 = 0 \\ 1, & \text{if } p_1(x) \leq 2\ell + bs \text{ and } x_1 = 1 \\ j, & \text{if } p_1(x) = 2\ell + bs + 2\ell_n + 1 + (j-2)(2\ell_n + 1) \text{ and } 2 \leq j < k \\ k + \frac{p_1(x) - (2\ell + bs + (k-1)(2\ell_n + 1))}{\ell_n + s + \ell_s + 1}, & \text{otherwise.} \end{cases}$$

It is easily verified that all statements made in the above proof for $k = 2$ colors remain correct if we consider the general case of $k \geq 3$ colors. Only the computation of the contributions $\Delta_{i(x)}(\text{BLOCK}_{i(x)}(x))$, equation (3), becomes a bit more tedious. However, all calculations are straightforward. We omit the details. □

# References

[AW09]  Gautham Anil and R. Paul Wiegand, *Black-box search by elimination of fitness functions*, Proc. of Foundations of Genetic Algorithms (FOGA'09), ACM, 2009, pp. 67–78.

[CCH96] Zhixiang Chen, Carlos Cunha, and Steven Homer, *Finding a hidden code by asking questions*, Proc. of the Second Annual International Conference on Computing and Combinatorics (COCOON'96), Springer, 1996, pp. 50–55.

[Chv83]  Vasek Chvátal, *Mastermind*, Combinatorica **3** (1983), 325–329.

[DJW06] Stefan Droste, Thomas Jansen, and Ingo Wegener, *Upper and lower bounds for randomized search heuristics in black-box optimization*, Theory of Computing Systems **39** (2006), 525–544.

[DW11]  Benjamin Doerr and Carola Winzen, *Memory-restricted black-box complexity*, ECCC TR11-092 (2011).

[ER63]   Paul Erdős and Alfréd Rényi, *On two problems of information theory*, Magyar Tud. Akad. Mat. Kutató Int. Közl. **8** (1963), 229–243.

[Goo09]  Michael T. Goodrich, *On the algorithmic complexity of the mastermind game with black-peg results*, Information Processing Letters **109** (2009), 675–678.

[Knu77]  Donald E. Knuth, *The computer as a master mind*, Journal of Recreational Mathematics **9** (1977), 1–5.

[SZ06]   Jeff Stuckman and Guo-Qiang Zhang, *Mastermind is NP-complete*, INFOCOMP Journal of Computer Science **5** (2006), 25–28.